\documentclass[superscriptaddress,citeautoscript,floatfix,longbibliography,bibnotes]{revtex4-2}
\usepackage{times,bm,siunitx,amsmath,amsfonts,amsthm,amssymb,amsbsy,braket,graphicx,hyperref}
\usepackage[utf8]{inputenc}
\usepackage[T1]{fontenc}
\usepackage[capitalize]{cleveref}
\usepackage{microtype}
\usepackage{times}
\newcommand{\ii}{\mathrm{i}}
\newcommand{\ee}{\mathrm{e}}
\newcommand{\dd}{\mathrm{d}}
\renewcommand{\Re}{\mathrm{Re}}
\renewcommand{\Im}{\mathrm{Im}}
\DeclareMathOperator{\sgn}{sgn}
\hypersetup{colorlinks=true,urlcolor=blue,linkcolor=blue,citecolor=red}

\begin{document}
\title{
Topological zero modes and bounded modes at smooth domain walls: 
Exact solutions and dualities
}
\author{Pasquale Marra}
\email{pasquale.marra@keio.jp}
\affiliation{
Graduate School of Mathematical Sciences, The University of Tokyo, 3-8-1 Komaba, Meguro, Tokyo, 153-8914, Japan
}
\affiliation{
Department of Physics, and Research and Education Center for Natural Sciences, Keio University, 4-1-1 Hiyoshi, Yokohama, Kanagawa, 223-8521, Japan}
\author{Angela Nigro}
\affiliation{
Dipartimento di Fisica ``E. R. Caianiello'', Università degli Studi di Salerno, Via Giovanni Paolo II, 132, 84084 Fisciano, Salerno, Italy
}
\date{\today}

\begin{abstract}
Topology describes global quantities invariant under continuous deformations, such as the number of elementary excitations at a phase boundary, without detailing specifics. 
Conversely, differential laws are needed to understand the physical properties of these excitations, such as their localization and spatial behavior. 
For instance, topology mandates the existence of solitonic zero-energy modes at the domain walls between topologically inequivalent phases in topological insulators and superconductors.
However, the spatial dependence of these modes is only known in the idealized (and unrealistic) case of a sharp domain wall.
Here, we find the analytical solutions of these zero-modes by assuming a smooth and exponentially-confined domain wall.
This allows us to characterize the zero-modes using a few length scales: the domain wall width, the exponential decay length, and oscillation wavelength. 
These quantities define distinct regimes: 
featureless modes with "no hair" at sharp domain walls, and nonfeatureless modes at smooth domain walls, respectively, with "short hair", i.e., featureless at long distances, and "long hair", i.e., nonfeatureless at all length scales.
We thus establish a universal relation between the bulk excitation gap, decay rate, and oscillation momentum of the zero modes, which quantifies the bulk-boundary correspondence in terms of experimentally measurable physical quantities. 
Additionally, we reveal an unexpected duality between topological zero modes and Shockley modes, unifying the understanding of topologically-protected and nontopological boundary modes.
These findings shed some new light on the localization properties of edge modes in topological insulators and Majorana zero modes in topological superconductors and on the differences and similarities between topological and nontopological zero modes in these systems. 
\end{abstract}


\maketitle

\section*{Introduction}

Topological laws are a special form of integral laws describing global quantities that are invariant under smooth and continuous deformations.
For instance, the total number of elementary excitations in a physical system, such as vortices in a fluid, defects in a crystal, boundary modes in quantum Hall insulators~\cite{thouless_quantized_1982}, or vortex modes in spin-triplet superconductors~\cite{volovik_fermion_1999,read_non-abelian_2009}, is a topological invariant regardless of the specific microscopic details and of the local differential laws governing the system.
Conversely, only differential equations can accurately describe the local physical properties of these excitations, such as their localization and spatial behavior.
This is evident in the case of the Jackiw-Rebbi equation in theoretical physics, describing Dirac fermions coupled to scalar fields~\cite{jackiw_solitons_1976}:
topological considerations alone mandate the existence of solitonic zero-energy modes localized around the topological defect, such as a domain wall.
However, to describe the spatial dependence of these modes, one still needs to find the analytical solutions of the differential equation:
In this specific case, the Jackiw-Rebbi equation, which is a first-order linear differential equation, is easily solvable by integration for any form of the scalar field considered~\cite{jackiw_solitons_1976}.

In topological insulators and superconductors in condensed matter~\cite{schnyder_classification_2009,hasan_colloquium:_2010,qi_topological_2011,shen_topological_2011,wehling_dirac_2014,shen_topological_2017}, the zero-energy solutions of a modified Jackiw-Rebbi equation correspond to boundary modes localized at the domain wall between topologically trivial and nontrivial phases~\cite{hatsugai_chern_1993,ryu_topological_2002,teo_topological_2010,shen_topological_2011,shen_topological_2017}.
This modified Jackiw-Rebbi equation, which is a second-order linear differential equation, describes nonrelativistic fermions in the presence of a superconducting pairing potential or semi-relativistic fermions with strong spin-orbit couplings.
Again, topological considerations mandate the existence of zero-energy modes localized around domain walls between topologically nonequivalent phases:
The number of boundary modes is equal to the absolute value of the difference between the topological invariants of the nonequivalent phases, a fundamental relation known as the bulk-boundary correspondence~\cite{hatsugai_chern_1993,ryu_topological_2002,teo_topological_2010}.
However, the general analytical solutions of this equation are not known, with the exception of the simple case where the domain wall is a sharp boundary, i.e., a Heaviside step function~\cite{shen_topological_2011,shen_topological_2017}.
Sharp boundaries, however, are only approximations: 
\emph{natura non facit saltus}, i.e., sharp boundaries do not exist since physical fields need to be continuous.

Here, we derive the analytical solutions of the modified Jackiw-Rebbi equation describing zero-energy modes at smooth domain walls between topological trivial and nontrivial phases.
To derive these solutions, we assume that the domain wall is exponentially confined with a finite localization length $w$.
The explicit analytical form of the wavefunction allows us to characterize the zero-modes in terms of a few length scales, i.e., the width of the domain wall $w$, the exponential decay length $\xi$, and the wavelength $\lambda$ describing the oscillations of the wavefunction. 
These quantities allow one to distinguish between qualitatively different regimes:
featureless zero modes localized at sharp domain walls $w\to0$ having "no hair" (in analogy with other featureless objects such as black holes);
nonfeatureless zero modes localized at smooth domain walls $\xi,\lambda>w>0$ having "short hair" (i.e., effectively featureless at long length scales);
nonfeatureless zero modes localized at smooth domain walls $w>\xi,\lambda>0$ having "long hair" (i.e., being nonfeatureless at all length scales).
Moreover, we show the existence of a universal relation between the bulk gap of the excitation spectra, the exponential decay rate $\mu=1/\xi$, and oscillation momentum $\kappa=2\pi/\lambda$ of the zero-modes, valid for both featureless and nonfeatureless modes.
This relation quantifies the bulk-boundary correspondence in terms of measurable physical quantities, opening the door to an accurate experimental characterization of zero-modes and the bulk-boundary correspondence.
Finally, we unveil a duality between the topological zero modes, solutions of the modified Jackiw-Rebbi equation, and the Shockley-like surface modes with exponential decay, solutions of the Schrödinger equation of a free particle in the presence of a smooth potential wall.
These findings shed new light on the connection between topological modes and nontopological surface modes, and on the localization properties of zero modes in topological insulators and superconductors.
Moreover, these findings contribute to understanding the differences and similarities between topologically protected boundary modes in topological superconductors, i.e., Majorana zero modes~\cite{read_non-abelian_2009,kitaev_unpaired_2001,alicea_new-directions_2012,leijnse_introduction_2012,das-sarma_majorana_2015,sato_majorana_2016,aguado_majorana_2017,laubscher_majorana_2021,marra_majorana_2022,tanaka_theory_2024}, 
and other zero-energy modes with seemingly no topological nature, such as trivial Andreev bound states~\cite{kells_near-zero-energy_2012,stanescu_disentangling_2013,liu_andreev_2017,hell_distinguishing_2018,liu_distinguishing_2018,penaranda_quantifying_2018,avila_non-hermitian_2019,vuik_reproducing_2019,yavilberg_differentiating_2019,prada_from_2020,pan_physical_2020,pan_crossover_2021,tian_distinguishing_2021,cayao_distinguishing_2021,liu_majorana_2021,marra_majorana/andreev_2022}.

\section*{Zero modes}

We consider the Jackiw-Rebbi equation~\cite{jackiw_solitons_1976} with an additional quadratic term in the momentum given by
\begin{equation}
	\left(\left(\eta p^2+ m(x)\right)\sigma_z
	+
 {2 v}(x)
 p\,\sigma_y
 \right)\Psi(x)=E \Psi(x),
\label{eq:H-pwaveC}
\end{equation}
where $\Psi(x)^\dag=(\psi(x)^\dag,\psi(x))$ is a spinor, $\sigma_{xyz}$ the Pauli matrices, $p=-{\ii}\partial$ the momentum operator, $\eta\ge0$, and with $ m(x), v(x)\in\mathbb{R}$.
This equation describes a Dirac fermion of mass $ m(x)$ and speed of light $2 v(x)$ if $\eta=0$.
In condensed matter, this equation describes a topological insulator with effective mass $1/2\eta$, chemical potential $\mu(x)=- m(x)$, and spin-orbit coupling $2 v(x)$,
or a spinless topological superconductor with effective mass $1/2\eta$, chemical potential $\mu(x)=- m(x)$, and $p$-wave superconducting pairing $\Delta_p(x)=2 v(x)$.
Here, we only focus on the zero modes $E=0$ which are normalizable and satisfy Dirichlet boundary conditions on infinite or semi-infinite intervals. 
These modes are necessarily eigenstates of $\sigma_x$ in the form
\begin{equation}\label{eq:spinor}
\Psi(x)\propto
\begin{pmatrix} 
1 \\ 
s
\end{pmatrix}
\varphi(x),
\quad
\sigma_x\Psi(x)=s\Psi(x)
,
\end{equation}
for $s=\pm1$ and with $\varphi(x)$ satisfying 
\begin{equation}\label{eq:diffeq}
\varphi''(x)
+{2s v}(x) \varphi'(x)
-
 m(x)
\varphi(x)
=0,
\end{equation}
where we absorbed $\eta$ into the definition of the fields $ v(x)\to  v(x)/\eta$ and $ m(x)\to  m(x)/\eta$.
We refer to the eigenvalue $s$ as the pseudospin.
Notice that the equation is real with real solutions and consequently the zero modes are Majorana modes analogous to the Majorana solutions of the Dirac equations.
Hence, the real eigenstates of the equation above are also eigenstates of the antiunitary complex conjugation operator $K$ and of the antiunitary operator $C=K\sigma_x$, which can be interpreted as the antiunitary time-reversal and charge conjugation (or particle-hole symmetry in condensed matter), respectively.
Notice that \cref{eq:H-pwaveC} can be unitary transformed via $ v(x) p \sigma_y  \to v(x) p \left( \sin{\theta}\, \sigma_y + \cos{\theta}\, \sigma_x \right)$, with $\theta$ an arbitrary gauge angle.
In the specific case of a superconductor, this corresponds to the gauge invariance with respect to the global phase of the superconductor.
Moreover, one can consider a more general equation where the gauge angle is position dependent $\theta\to\theta(x)$ giving $ v(x) p \sigma_y  \to v(x) p \left( \sin{\theta(x)}\, \sigma_y + \cos{\theta(x)}\, \sigma_x \right)$.
In this case, however, the solutions cannot be written in terms of eigenstates of a single Pauli matrix for all values of the position $x$ as in \cref{eq:spinor}.
We find numerical evidence that for $\theta'(x)\neq0$, there are no zero energy solutions of the Jackiw-Rebbi equation unless in the limiting case $m(x)=0$.
This limiting case corresponds to the closing of the bulk gap of the topological insulator or topological superconductor, yielding a continuous spectra around zero energy, and no localized zero-modes.

For uniform fields $ m(x)= m$, $ v(x)= v$, one can define a topological invariant $W\in\mathbb{Z}_2$ characterizing the Hamiltonian in \cref{eq:H-pwaveC}, which is $W=\sgn(v)$ for $ m\le0$ and $W=0$ otherwise~\cite{kitaev_unpaired_2001}.
Due to the bulk-boundary correspondence~\cite{hatsugai_chern_1993,ryu_topological_2002,teo_topological_2010}, zero energy modes localize at the boundary between topologically trivial $W=0$ and nontrivial regions $W\neq0$, or at the boundary between topologically nontrivial phases with different topological invariants.
The number of zero energy modes localized at the boundary is equal to the difference between the topological invariant on the right and on the left of the boundary $\Delta W=|W_L-W_R|$.

We assume that the fields approach their limit values on the left and right exponentially in the system size
$| m(x\to\mp\infty)- m_{L,R}|\propto{\ee}^{- |x|/w}$,
$|v(x\to\mp\infty)-v_{L,R}|\propto{\ee}^{- |x|/w}$,
where $w>0$ is a characteristic length describing the spatial variations of fields, i.e., the width of the smooth domain wall localized at the origin $x=0$.
By mapping the whole real line $-\infty<x<\infty$ into the finite segment $0<y<1$ with the substitution $y(x)=(1+\tanh{(x/2w)})/2$, one obtains
\begin{equation}\label{eq:diffeqtanh}
\varphi''(y)
+\frac{1 - 2y+2w s v(y)}{y(1-y)}
\varphi'(y)
-\frac{
w^2 m(y)
}{y^2(1-y)^2} 
\varphi(y)
=0,
\end{equation}
which presents two regular singular points at $y=0,1$.
Hence, assuming a solution $\varphi(y)=y^{w\alpha_L}(1-y)^{w{\alpha_R}} F(y)$ by solving indicial equations for the exponents at the singular points we find
the exponents
$\alpha_L^\pm=-s v_L\pm q_{L}$, 
$\alpha_R^\pm=s v_R\pm q_{R}$,
where $ q_{L,R}=\sqrt{ v_{L,R}^2+ m_{L,R}}$.
Hence, for a given pseudospin $s=\pm1$, the general solution of the modified Jackiw-Rebbi equation is given by a linear combination $\sum_i A_i\varphi_i^s(x)$ where
\begin{align}
 \varphi_{1,2}^s(x)=&
 \frac{{\ee}^{({ \alpha_L^\pm}-{\alpha_R^\mp})x/2} }
 {\left(2\cosh{\left(\frac{x}{2w}\right)}\right)^{{w(\alpha_L^\pm}+{\alpha_R^\mp})}}
 F_{1,2}(x)
,
\label{eq:2generalsolutionsMAIN}
\end{align}
which depend only on the values of the fields at large distances $|x|\gg w$, and where the functions $F_{1,2}$ are bounded or diverge polynomially for $x\to\pm\infty$.
Asymptotically, these solutions give 
$\varphi_{1,2}^s(x\to-\infty)\propto{\ee}^{- \alpha_L^\pm |x|}={\ee}^{- \mu_L^\pm |x|}{\ee}^{\mp {\ii}\kappa_L |x|}$,
$\varphi_{1,2}^s(x\to+\infty)\propto{\ee}^{- \alpha_R^\mp |x|}={\ee}^{- \mu_R^\mp |x|}{\ee}^{\pm {\ii}\kappa_R |x|}$,
where $\mu_{L,R}^\pm=1/\xi_{L,R}^\pm=\Re(\alpha_{L,R}^\pm)$ are the decay rates and decay lengths,
while $\kappa_{L,R}=2\pi/\lambda_{L,R}=\Im( q_{L,R})$ the momentum and wavelengths responsible for oscillatory behaviors.
For $v_{L,R}^2+ m_{L,R}\ge0$ one has
$\kappa_{L,R}=0$,
with
$\mu_{L,R}^+=\mu_{L,R}^-$ in the limiting case $v_{L,R}^2+ m_{L,R}=0$.
Conversely, for $ v_{L,R}^2+ m_{L,R}<0$ one has that
$\mu_{L}^+=\mu_{L}^-=- s v_{L}$,
$\mu_{R}^+=\mu_{R}^-= s v_{R}$, and
$\kappa_{L,R}=\Im( q_{L,R})\neq0$.

For simplicity, we assumed that the characteristic lengths describing the variations of the two fields coincide.
Nevertheless, \cref{eq:diffeqtanh,eq:2generalsolutionsMAIN} remain valid even in the assumption that the two decay lengths are different, i.e., 
$| m(x\to\mp\infty)- m_{L,R}|\propto{\ee}^{- |x|/w_1}$,
$| v(x\to\mp\infty)- v_{L,R}|\propto{\ee}^{- |x|/w_2}$,
by defining $w=\max(w_1,w_2)$, since the longer decay length is dominant at large distances ${\ee}^{- |x|/w}\ge{\ee}^{- |x|/w_1}$ and ${\ee}^{- |x|/w}\ge{\ee}^{- |x|/w_2}$.
However, to proceed further and obtain closed-form analytical solutions, we must assume that the two characteristic lengths coincide. 
This is justified since we are interested in physical systems with smooth transitions between two distinct phases with distinct values of the fields $ v(x)$ and $ m(x)$, such that the transition is confined in a small region of width $w$.
In this case, in first approximation, the variations of the fields can be assumed to converge exponentially to their values at $\pm\infty$ with same decay length $w$.
This assumption is further justified in the specific scenario where $ v(x)$ and $ m(x)$ describe the superconducting parameter and the chemical potential of a topological superconductor.
In the case of intrinsic superconducting order, the gap equation defining the superconducting order parameter depends directly on the Fermi level, which is directly affected by the chemical potential.
Arguably, in this case, an exponential dependence of the chemical potential induces a similar exponential dependence in the order parameter.
Moreover, assuming that 
$ v(x)$ and $ m(x)$ can be expanded in powers of $y(x)=\left(1+\tanh{(x/2w)}\right)/2$ up to the first and second order, respectively, i.e., as 
$ v(x)= v_0+ v_1 y(x)$, 
and
$ m(x)= m_0+ m_1 y(x)+ m_2 y(x)^2$,
where 
$ v_0= v_L$, $ v_1= v_R- v_L$,
$ m_0= m_L$, $ m_1+ m_2= m_R- m_L$,
and $ m_2= 2( m_L +  m_R)-4  m_D$,
with $ m_D= m(0)$,
we find an explicit general solution in terms of hypergeometric functions.
For each set of exponents $({\alpha_L},{\alpha_R})=(\alpha_L^\pm,\alpha_R^\pm)$ the equation becomes
\begin{equation}
F''(y)
+
\frac{c-(a+b+1) y}{y(1-y)} 
F'(y)
-
\frac{a b}{y(1-y)}
F(y)
=0,
\label{eq:diffeqfhg}
\end{equation}
which is the hypergeometric equation with
$a,b=w{\alpha_L}+w{\alpha_R} -w s v_1
+\frac12
\pm
\frac12\sqrt{
\left(
2ws v_1
-1\right)^2
+4w^2 m_2
}
$ and
$c=
1+2ws v_L+2w{\alpha_L}$,
The solutions of the hypergeometric equation can be chosen as two linearly independent solutions out of the Kummer's 24 solutions of the hypergeometric equation, e.g., taking ${\,}_2F_1\left(a,b, c,y\right)$, and ${\,}_2F_1\left(a,b, a+b-c+1,1-y\right)$.
Hence, for a given pseudospin $s=\pm1$, we found that
\begin{align}
 F_{1,2}(x)=&
 {{\,}_2F_1\left(a_{1,2},b_{1,2},c_{1,2},\tfrac{1}{{\ee}^{\mp x/w}+1}\right)}
 ,
\label{eq:2generalsolutionshyperMAIN}
\end{align}
where
${\,}_2F_1(a,b,c,y)$ is the hypergeometric function with
\begin{subequations}\label{eq:ab1ab2MAIN}
\begin{align} 
a_{1,2}&=\pm w\left({ q_L}-{ q_R}\right) +\tfrac12+\tfrac12{\sqrt{\left(2ws v_1- 1\right)^2+4w^2 m_2}},
\\
b_{1,2}&=\pm w\left({ q_L}-{ q_R}\right) +\tfrac12-\tfrac12{\sqrt{\left(2ws v_1- 1\right)^2+4w^2 m_2}},
\\
c_{1,2}&=1+2w  q_{L,R},
\end{align}
\end{subequations}
assuming $c_{1,2}-a_{1,2},c_{1,2}-b_{1,2}\notin\mathbb{Z}_\leq$ (we denote $\mathbb{Z}_\leq$ the set of nonpositive integers).

The solutions $\varphi_{1,2}^s(x)$ simplify in the cases where $| v_{L,R}|= v$ or $ m_{L,R}= m$ or if the fields are uniform, as shown in \cref{tab:specialcases}.
The functions $F_{1,2}$ are bounded or diverge polynomially for all choices of the parameters. 
Hence, the exponents $\alpha_L^\pm$ and $\alpha_R^\pm$ uniquely determine the asymptotic behavior of the solutions and whether there exist particular solutions that satisfy the boundary conditions on a given interval, i.e., whether zero modes exist on a finite, semi-infinite, or infinite system.
These exponents depend uniquely on the values of the fields at large distances $ m_{L,R}$ and $ v_{L,R}$.
The parameters $a_{1,2}$, $b_{1,2}$, and $c_{1,2}$ depend as well on the values of the fields at large distances $ m_{L,R}$ and $ v_{L,R}$.
However, the two parameters $a_{1,2}$ and $b_{1,2}$ also depend on the values of the field $ m(x)$ at the domain wall via $ m_2$, since $ m_2= 2( m_L +  m_R)-4  m_D$ where $ m_D= m(x=0)$ is the value of the field $ m(x)$ at the center of the domain wall $x=0$.
Note that the value of $m_2$ gives a measure of the variations of the field around the domain wall, since $m_2/4$ equals to the difference between the value of the field at $x=0$ and the average of the field at large distances $x\to\pm\infty$.

\begin{table*}[t]
 \centering
 \includegraphics[width=\textwidth]{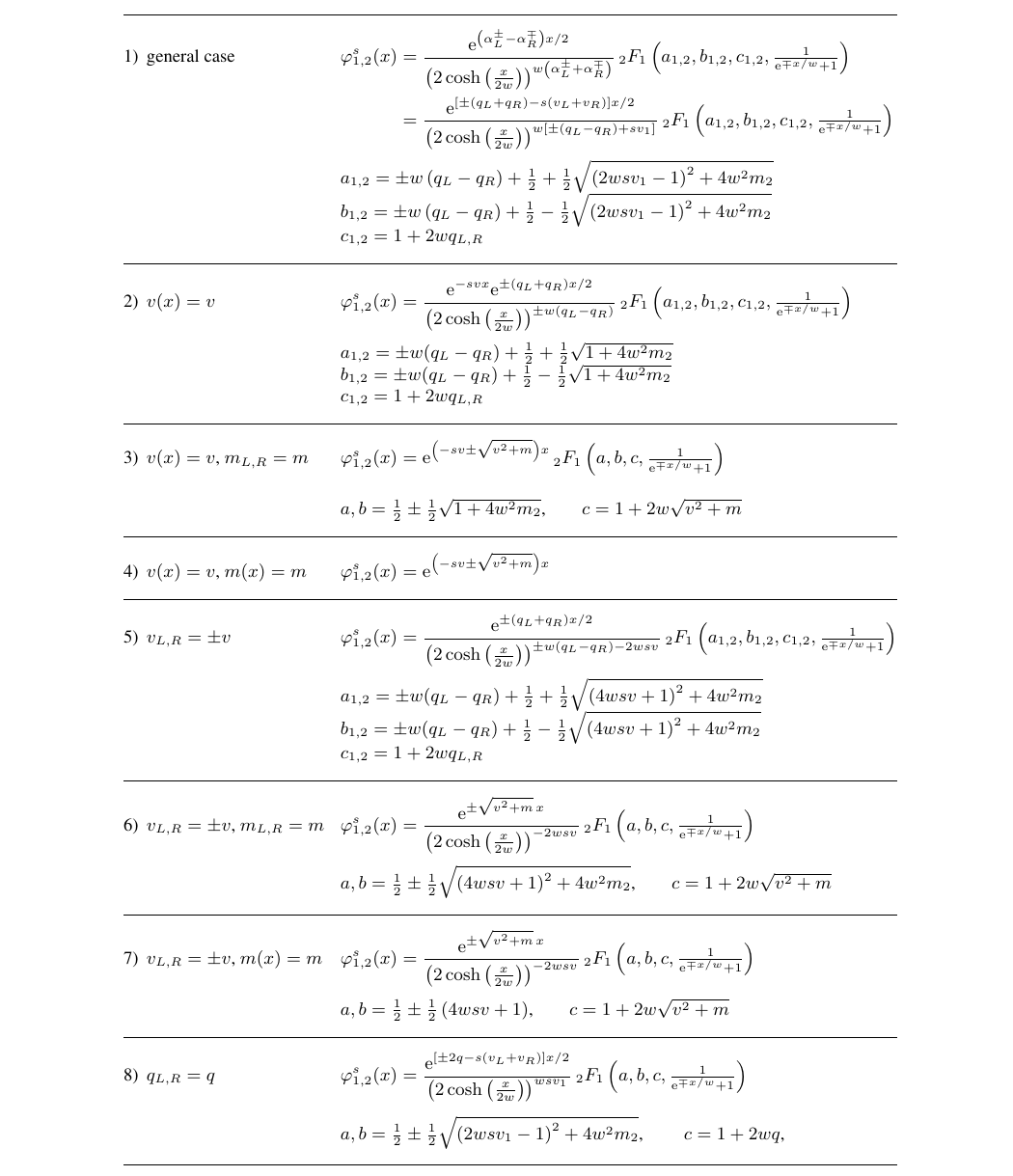} 
\caption{
The general solutions of the modified Jackiw-Rebbi equation in terms of hypergeometric functions for 
$ m(x)= m_0+ m_1 y(x)+ m_2 y(x)^2$ and
$ v(x)= v_0+ v_1 y(x)$,
and for some special cases, i.e., where $ v(x)= v_{L,R}= v$ is constant or a 
symmetric S-shaped curve 
($ v_0= v$ and $ v_1=-2 v$) 
with $ v_{L,R}=\pm v$,
and where $ m(x)= m$ is constant
or a symmetric Pöschl–Teller potential 
($ m_0= m$ and $ m_1+ m_2=0$), 
with $ m_{L,R}= m$,
and more generally, the case where $ q_L= q_R$.
If $ q_{L,R}= q$, then one gets $a_{1,2}=a$ and $b_{1,2}=b$ with $a+b=1$ in rows 3, 6, 7, and 8:
In this case one has also
 ${{\,}_2F_1\left(a,b,c,\frac{1}{{\ee}^{\mp x/w}+1}\right)}
=\Gamma(c)
{\ee}^{\pm  q x} P^{1-c}_{-a}\left(\tanh\left(\mp\frac{x}{2w}\right)\right)
$.
The case on row 4 coincides with the well-known solution describing the end mode of a uniform one-dimensional topological insulator, or the Majorana end mode of a uniform one-dimensional spinless topological superconductor~\cite{shen_topological_2011}.
These are the general solutions for all choices of the parameters, with the exceptions of the limiting cases where either
$c_{1}-a_{1}=c_{2}-a_{2}\in\mathbb{Z}_\leq$ or $c_{1}-b_{1}=c_{2}-b_{2}\in\mathbb{Z}_\leq$.
}
 \label{tab:specialcases}
\end{table*}

The values of the fields at large distances and the boundary conditions univocally determine the existence and number of particular solutions and their pseudospin $s$ on infinite and semi-infinite intervals.
In particular, to ascertain the existence, number, and asymptotic properties of the solutions, we define the topological invariants 
\begin{equation}\label{eq:TI}
W_{L,R}=
\begin{cases}
\sgn( v_{L,R}), & \text{ if }  m_{L,R}\le0,
\\
0 & \text{ if }  m_{L,R}>0.
\end{cases}
\end{equation}
These topological invariants depend only on the values of the fields $ m_{L,R}$ and $ v_{L,R}$ at large distances $|x|>w$.
The phase on the left $x\ll -w$ is topologically nontrivial if $ m_L\le0$ and $ v_L\neq0$ with topological invariant equal to $W_L=\sgn( v_L)$ and trivial otherwise.
Analogously, the phase on the right $x\gg w$ is nontrivial if $ m_R\le0$ and $ v_R\neq0$  with $W_R=\sgn( v_R)$ and trivial otherwise (see Refs.~\cite{kitaev_unpaired_2001,shen_topological_2011}).

Zero modes on the interval $-\infty<x<\infty$ satisfying the boundary conditions $\varphi(\pm\infty)=0$ are given by the modes $\varphi_{1,2}^s(x)$ in \cref{eq:2generalsolutionsMAIN} with $\Re(\alpha_L^\pm)>0$ and $\Re(\alpha_R^\pm)>0$.
As one would expect from the bulk-boundary correspondence~\cite{hatsugai_chern_1993,ryu_topological_2002,teo_topological_2010}, there are two independent modes if $W_{L}W_R=-1$, one independent mode if $W_{L}=\pm1$ and $W_R=0$ or if $W_{R}=\pm1$ and $W_L=0$.
Indeed, for $\sgn( m_{L,R})=-1$ and $W_LW_R=-1$, taking $s=-W_L=W_R$ gives $\Re(\alpha_L^\pm),\Re(\alpha_R^\pm)>0$:
Hence, both $\varphi_{1,2}^s(x)$ decay exponentially to zero for $x\to\pm\infty$,
and consequently the particular solution satisfying the boundary conditions is any linear combination $\varphi(x)=\sum_i A_i\varphi_i^s(x)$. 
The asymptotic behavior of the modes $\varphi_1^s$ and $\varphi_2^s$ are determined 
on the left side by the exponents 
$\alpha_{L}^+=-s v_{L}+ q_{L}$ 
(characteristic lengths $\xi_L^+$, $\lambda_L^+$)
and 
$\alpha_{L}^-=-s v_{L}- q_{L}$
(characteristic lengths $\xi_L^-$, $\lambda_L^-$),
respectively,
and on the right side by the exponents 
$\alpha_{R}^-=s v_{R}- q_{R}$ 
(characteristic lengths $\xi_R^-$, $\lambda_R^-$)
and
$\alpha_{R}^+=s v_{R}+ q_{R}$
(characteristic lengths $\xi_R^+$, $\lambda_R^+$),
respectively.

For $\sgn( m_L)=1$, $\sgn( m_R)=-1$ (i.e., $W_L=0$ and $|W_R|=1$) and in the limiting case 
$\sgn( m_L)=0$, $\sgn( m_R)=-1$ with $W_LW_R=-1$,
choosing $s=W_R$ gives $\Re(\alpha_L^+)>0>\Re(\alpha_L^-)$ and $\Re(\alpha_R^\pm)>0$:
Hence, $\varphi_1^s(x)$ decays to zero for $x\to\pm\infty$ while $\varphi_2^s(x)$ diverges for $x\to-\infty$, and thus there is a single mode satisfying the boundary conditions, which is $\varphi(x)=A_1\varphi_1^s(x)$.
The asymptotic behavior is determined 
on the left by the exponents 
$\alpha_{L}^+=-s v_{L}+ q_{L}$ (characteristic lengths $\xi_L^+$, $\lambda_L^+$)
and on the right by
$\alpha_{R}^-=s v_{R}- q_{R}$ (characteristic lengths $\xi_R^-$, $\lambda_R^-$).

Analogous arguments hold for $\sgn( m_R)=1$, $\sgn( m_L)=-1$ (i.e., $W_R=0$ and $|W_L|=1$), and for the limiting case 
$\sgn( m_R)=0$, $\sgn( m_L)=-1$ with $W_LW_R=-1$,
taking $s=-W_L$, giving a single mode satisfying the boundary conditions $\varphi(x)=A_2\varphi_2^s(x)$, with asymptotic behavior determined 
on the left by the exponents 
$\alpha_{L}^-=-s v_{L}- q_{L}$ (characteristic lengths $\xi_L^-$, $\lambda_L^-$)
and on the right by 
$\alpha_{R}^+=s v_{R}+ q_{R}$ (characteristic lengths $\xi_R^+$, $\lambda_R^+$).

Zero modes on the interval $0\le x<\infty$ satisfying the boundary conditions $\varphi(0)=\varphi(\infty)=0$ are given by linear combinations of $\varphi_{1,2}^s(x)$ in \cref{eq:2generalsolutionsMAIN} with $\Re(\alpha_R^\pm)>0$.
As one would expect from the bulk-boundary correspondence~\cite{hatsugai_chern_1993,ryu_topological_2002,teo_topological_2010}, there is one independent mode if $W_R\neq1$.
For $\sgn( m_R)=-1$ and $|W_R|=1$, choosing $s=W_R$ gives $\Re(\alpha_R^\pm)>0$, and therefore 
both $\varphi_{1,2}^s(x)$ decays to zero for $x\to\infty$, and thus one can obtain a single particular solution satisfying the boundary conditions, which is the linear combination $\varphi(x)=\sum_i A_i\varphi_i^s(x)$ with $A_{1,2}$ chosen such that $\varphi(0)=0$.
The asymptotic behavior is determined by the 
exponent $\alpha_{R}^-=s v_{R}- q_{R}$ 
(characteristic lengths $\xi_R^-$, $\lambda_R^-$).
This is because, even though this mode is a linear combination of two exponentially decaying solutions with decaying lengths $\xi_R^\pm$, one has that $\xi_R^-\ge \xi_R^+$, which means that the mode $\varphi_1^s$ with decaying length $\xi_R^-$ dominates at large distances.
In the special cases where $| v_{L,R}|= v$ and $ m_R= m_L$, one has ${A_1}/{A_2} = -1$.
Zero modes on the interval $-\infty<x\le 0$ can be obtained similarly.

The oscillatory behavior of the modes is determined by the quantities $ v_{L,R}^2+ m_{L,R}$:
One has that
$ v_{L,R}^2+ m_{L,R}\ge0$ gives $\kappa_{L,R}=0$, corresponding to exponential decay for $x\to\mp\infty$, respectively;
Conversely, $ v_{L,R}^2+ m_{L,R}<0$ gives $\kappa_{L,R}\neq0$,  corresponding to exponentially damped oscillations for $x\to\mp\infty$, respectively, in summary
\begin{equation}
\begin{cases}
 v_{L,R}^2+ m_{L,R}\ge0 \rightarrow \text{exp.~decay}
\\
 v_{L,R}^2+ m_{L,R}<0 \rightarrow \text{exp.~damped oscill.}
\end{cases}
\text{ for } x\to\mp\infty.
\end{equation}

Note that, since the modified Jackiw-Rebbi equation is real, its solutions are real for some choices of $A_{1,2}$.
The existence and number of particular solutions on the infinite and semi-infinite intervals $-\infty<x<\infty$ and $0\le x<\infty$ are summarized in \cref{tab:solintervals}.

\begin{table*}[t]
 \centering
 \includegraphics[width=\textwidth]{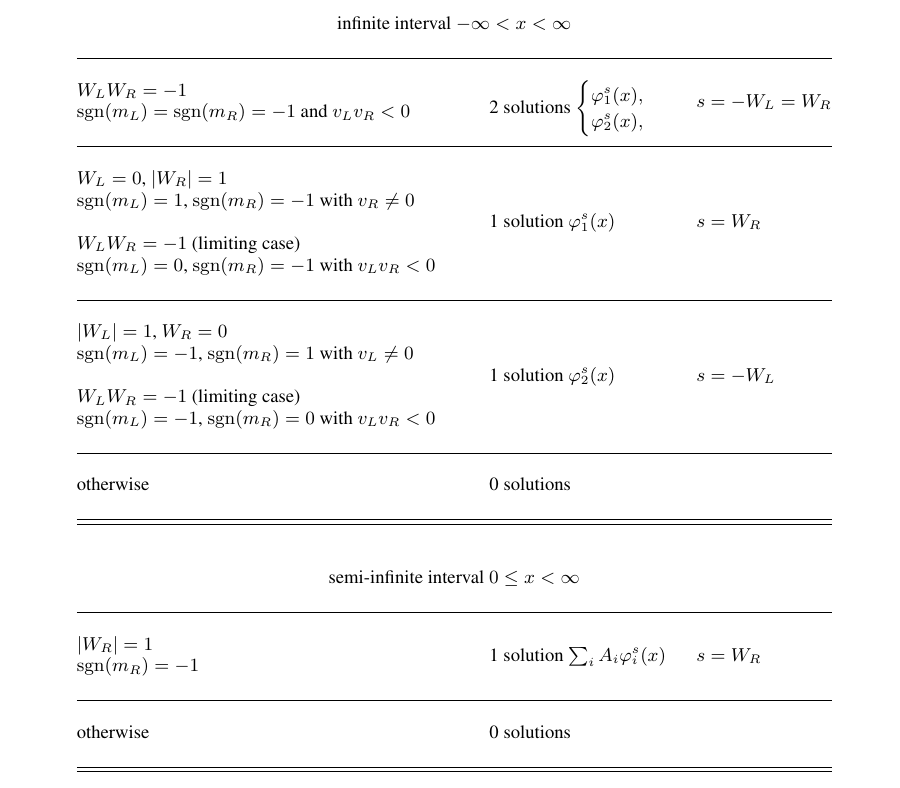}  
\caption{
The number of modes and pseudospin $s$ satisfying the boundary conditions 
$\varphi(\pm\infty)=0$ on the infinite interval $-\infty<x<\infty$ and
$\varphi(0)=\varphi(\infty)=0$ on the semi-infinite interval $0\le x<\infty$ 
depending on the asymptotic values 
$ m_{L,R}, v_{L,R}$, and $W_{L,R}$.
The modes show exponential decay if $ v_L^2+ m_L\ge0$ or exponentially damped oscillations if $ v_L^2+ m_L<0$.
The asymptotic behavior of the solutions is completely determined by the decay rates and decay lengths $\mu_{L,R}^\pm=1/\xi_{L,R}^\pm=\Re(\alpha_{L,R}^\pm)$ and the momenta and wavelengths $\kappa_{L,R}=2\pi/\lambda_{L,R}=|\Im(\alpha_{L,R}^\pm)|$ responsible for oscillatory behaviors.
Furthermore, 
$ v_{L,R}^2+ m_{L,R}\ge0$ yields $\kappa_{L,R}=0$,
and the limiting case $ v_{L,R}^2+ m_{L,R}=0$ yields $\mu_{L,R}^+=\mu_{L,R}^-$ and $\kappa_{L,R}=0$.
Also, 
$ v_{L,R}^2+ m_{L,R}<0$ yields
$\mu_{L}^+=\mu_{L}^-=- s v_{L}$,
$\mu_{R}^+=\mu_{R}^-= s v_{R}$,
and
$\kappa_{L,R}\neq0$.
}
 \label{tab:solintervals}
\end{table*}

\section*{Length scales and hairstyles}

Therefore, the existence and number of zero modes and their pseudospin in the infinite interval $-\infty<x<\infty$ only depend on the topological invariants $W_{L,R}$, i.e., on the values of the fields $ m_{L,R}$ and $ v_{L,R}$ at large distances $|x|>w$.
These modes are localized at the smooth domain wall, and their number coincides with the difference between the topological invariants on the left and right sides $|W_R-W_L|$, as expected from the bulk-boundary correspondence~\cite{hatsugai_chern_1993,ryu_topological_2002,teo_topological_2010}.
Analogously, the existence of a zero mode in the semi-infinite interval $0\le x<\infty$ only depends on the topological invariant $W_R$, i.e., on the values of the fields $ m_R$ and $ v_R$ at $x>w$, with its pseudospin determined only by $W_R$.
This mode is localized at an infinite wall at $x=0$, when the topological invariant on the right side is nontrivial $|W_R|=1$, as expected from the bulk-boundary correspondence.

Furthermore, at large distances $x/w\to\pm\infty$, the asymptotic behavior of each mode is determined only by the associated decay rates $\mu_{L,R}^\pm$ and momenta $\kappa_{L,R}$ or, alternatively, by the characteristic decay lengths $\xi_{L,R}^\pm$ and oscillation wavelengths $\lambda_{L,R}$ for each asymptote and for each mode, describing the exponential decay or exponentially damped oscillations, which ultimately depend only on the values of the fields $ m_{L,R}$ and $ v_{L,R}$ at large distances $|x|\gg w$.
On the left asymptote, the case $ v_{L}^2+ m_{L}\ge0$ mandates $\kappa_{L}=0$ (i.e., $\lambda_{L}=\infty$) corresponding to exponential decay (without oscillations), while the case $ v_{L}^2+ m_{L}<0$ mandates $\kappa_{L}\neq0$ (i.e., $\lambda_{L}=2\pi/\kappa_L$) corresponding to exponentially damped oscillations.
Moreover, one has that $\mu_L^+=\mu_L^-$ if $ v_{L}^2+ m_{L}\le0$.
Analogous statements hold for the right asymptote.

At short distances $|x|\lesssim w$ instead, the modes wavefunctions cannot be fully described by the characteristic lengths $\xi_{L,R}^\pm$, $\lambda_{L,R}$ (or by the complex numbers $\alpha_{L,R}^\pm$).
This is because, at short distances, the zero modes are not only determined by the values of the fields at the left and right asymptotes $ m_{L,R}$ and $ v_{L,R}$, but also by the width of the smooth domain wall $w$, and on the details of the spatial dependences of the fields.
Indeed, the zero modes in \cref{eq:2generalsolutionsMAIN} depend on the values $ m_D$, which is not fully determined by the asymptotic values $ m_{L,R}$.

We notice that the limit $w\to0$ describes the well-studied case of zero modes localized at sharp phase boundaries, i.e., a domain wall where the fields change abruptly.
These modes localized at a phase boundary at $x=0$ are only characterized by the characteristic lengths $\xi_R$, $\lambda_R$ or alternatively, by one complex number $\alpha_R$ for $x>0$ and another set of characteristic lengths $\xi_L$, $\lambda_L$ or alternatively, by one complex number $\alpha_L$ for $x<0$.
Hence, zero modes localized at sharp phase boundaries are featureless objects with "no hair" (in analogy with the usual expression used for other featureless objects such as black holes).
Crucially, these objects remain featureless at all length scales.

On the contrary, zero modes localized at a smooth domain wall have "hair" in the sense that they are nonfeatureless, i.e., cannot be described entirely by a small set of numbers, at least not at short length scales.
Indeed, in the case $w<\xi,\lambda$, modes localized at a smooth domain wall appear featureless at large distances $|x|>\xi,\lambda>w$ but not at short distances $|x|<w$.
As long as they are looked at from a large distance $|x|>\xi,\lambda>w$, these modes are effectively hairless, i.e., they can be treated as effectively featureless modes at long length scales, being as good as zero modes localized at sharp phase boundaries, and being characterized only by a few numbers ($\xi,\lambda$, or $\alpha$).
We refer to this case as zero modes with "short hair".
Dramatically different is the case where $w>\xi,\lambda$.
In this case, the features of the wavefunctions cannot be neglected since the length scales associated with these features are larger than the decaying lengths of the modes.
Hence, these wavefunctions are nonfeatureless at all length scales, i.e., cannot be described only by a set of few numbers $\xi,\lambda$, or $\alpha$ at any scale.
We refer to this case as zero modes with "long hair".
Notice also that the property of being a short-hair $w<\xi,\lambda$ or long-hair $w>\xi,\lambda$ mode depends on whether we look at the mode from the left side (i.e., $x<0$) or from the right side (i.e., $x>0$ for a smooth domain wall at $x=0$).
We notice that, even though their topological nature is not apparent at all length scales, the existence of the short-hair $w<\xi,\lambda$ and long-hair $w>\xi,\lambda$ modes at smooth domain walls is still a direct consequence of the bulk-boundary correspondence.
In the context of one-dimensional topological superconductors, featureless, hairless zero modes correspond to textbook examples of Majorana zero modes localized at a sharp boundary between a topologically nontrivial phase and a trivial phase (e.g., the vacuum), whereas nonfeatureless zero modes in the short-hair or long-hair regimes correspond to different incarnations of quasi-Majorana modes which appear at smooth boundaries~\cite{kells_near-zero-energy_2012,stanescu_disentangling_2013,liu_andreev_2017,hell_distinguishing_2018,liu_distinguishing_2018,penaranda_quantifying_2018,avila_non-hermitian_2019,vuik_reproducing_2019,yavilberg_differentiating_2019,prada_from_2020,pan_physical_2020,pan_crossover_2021,tian_distinguishing_2021,cayao_distinguishing_2021,liu_majorana_2021,marra_majorana/andreev_2022}.

\begin{figure}[tbp]
 \centering
 \includegraphics[width=\textwidth]{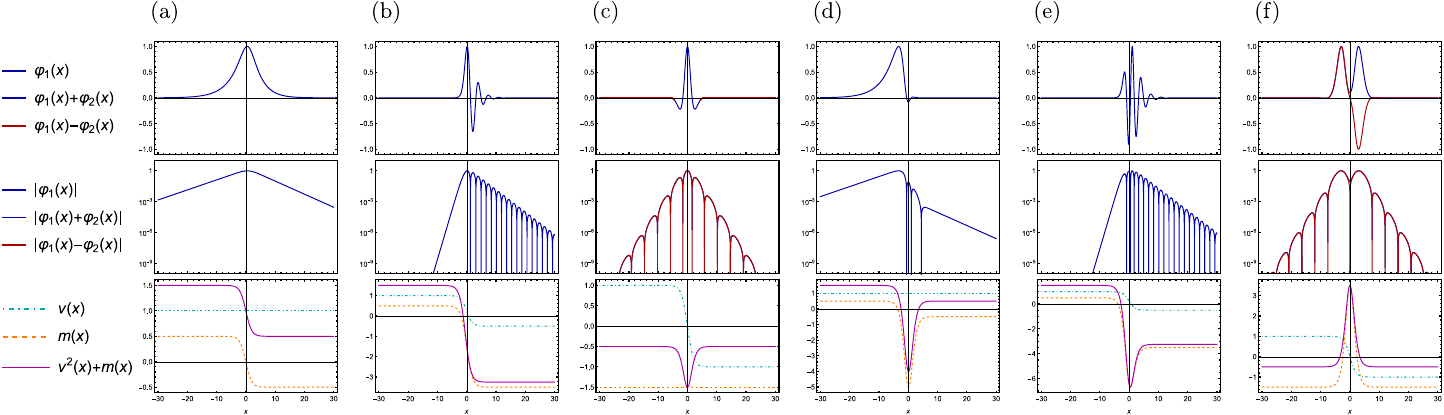} 
\caption{
Zero modes on the interval $-\infty<x<\infty$ with a smooth domain wall described by the fields $ m(x)$ and $ v(x)$ with a width smaller than or comparable with the characteristic decay lengths $w\lesssim\xi$.
Upper panels show the wavefunction, middle panels show the logarithm of the absolute value of the wavefunction, and lower panels show the fields $ v(x)$, $ m(x)$, and $ v(x)^2+ m(x)$.
Different columns correspond to different choices of the parameters $ m_{0,1,2}$ and $ v_{1,2}$:
(a)
single mode with exponential decay on both sides 
of a smooth domain wall with S-shaped field $ m(x)$ and constant $ v(x)$;
(b) 
single mode with exponentially damped oscillations on the right side and exponential decay on the left side
of a smooth domain wall with S-shaped fields $ m(x)$ and $ v(x)$;
(c) 
two modes with exponentially damped oscillations on both sides
of a smooth domain wall with S-shaped field $ v(x)$ and constant $ m(x)$
(we consider the two linear combinations $\varphi_1\pm\varphi_2$ giving real wavefunctions);
(d)
single mode with exponential decay on both sides 
of a smooth domain wall with asymmetric Pöschl–Teller field $ m(x)$ and constant $ v(x)$;
(e) 
single mode with exponentially damped oscillations on the right side and exponential decay on the left side
of a smooth domain wall with asymmetric Pöschl–Teller field $ m(x)$ and S-shaped field $ v(x)$;
(f) 
two modes with exponentially damped oscillations on both sides
of a smooth domain wall with symmetric Pöschl–Teller field $ m(x)$ and S-shaped field $ v(x)$
(linear combinations $\varphi_1\pm\varphi_2$).
}
 \label{fig:infinfshort}
~\\
 \includegraphics[width=\textwidth]{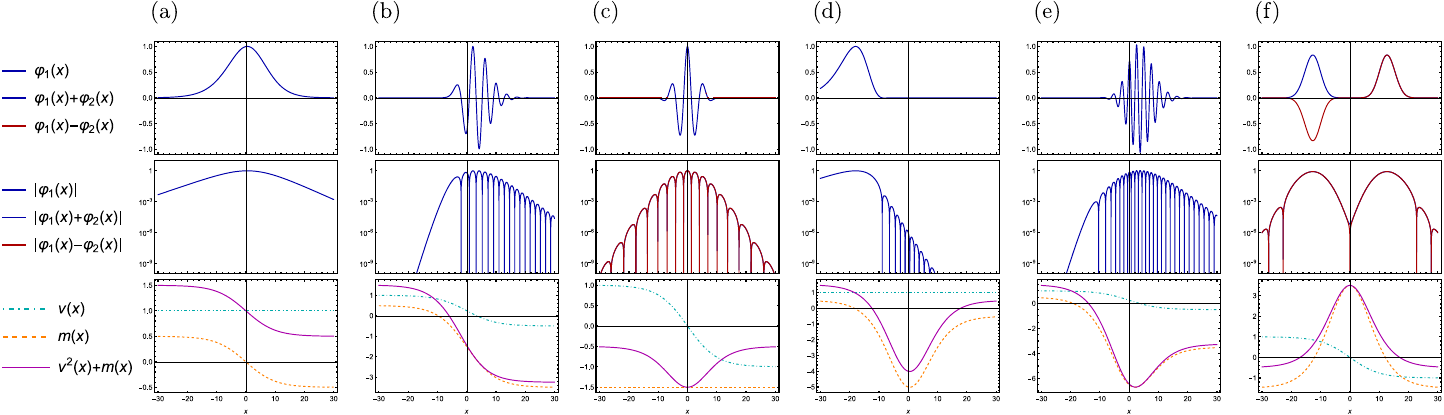} 
 \caption{
Same as in \cref{fig:infinfshort}, but with a smooth domain wall with a width larger than the characteristic decay lengths $w>\xi$.
}
 \label{fig:infinflong}
\end{figure}

The analytic solutions we found describe zero modes at smooth domain walls in all regimes, i.e., featureless and nonfeatureless cases with short-hair $w<\xi$ and long-hair $w>\xi$.
Moreover, the long-distance asymptotic behavior for $x\gg w$ is completely determined by the exponents $\mu$ and $\kappa$, and by the signs of the quantities $ m_{L,R}$ and $ v_{L,R}^2+ m_{L,R}$.
However, to gain a physical intuition on the localization of the zero modes at short distances $x\lesssim w$, one can analyze the spatial dependence of the quantities $ m(x)$ and $ v(x)^2+ m(x)$ and define the local topological invariant 
\begin{equation}\label{eq:TIRS}
W(x)=
\begin{cases}
\sgn( v(x)) & \text{ if }  m(x)\le0,
\\
0 & \text{ if }  m(x)>0,
\end{cases}
\end{equation}
which generalizes $W_{L,R}$ to the whole interval $-\infty<x<\infty$. 

We have seen that zero modes on the interval $-\infty<x<\infty$ appear when the topological invariant at long distances $W_{L,R}=W(x\to\pm\infty)$ differ on the right and left asymptotes $W_L\neq W_R$.
This mandates that the local topological invariant $W(x)$ must change value an odd number of times if $|W_L-W_R|=1$ or an even number of times if $|W_L-W_R|=2$ somewhere in the interval $-\infty<x<\infty$.
In particular, if the fields $ m(x)$ and $ v(x)$ can be expanded up the second order and up to the first order in $y(x)$, respectively, then $W(x)$ can change at most two times on the interval $-\infty<x<\infty$, i.e.,  one time if $|W_L-W_R|=1$ or two times if $|W_L-W_R|=2$.
These points $x_i^*$ must be $|x_i^*|\lesssim w$, since the spatial variations of the fields are confined within the smooth domain wall $-w\lesssim x\lesssim w$.
In the case where $w\gg\xi$, the distance between these points may be larger than the localization length $\xi$.
It is clear from the bulk-boundary correspondence that, in this case, zero modes will localize in correspondence of these points $x\approx x_i^*$. 

A similar argument can be applied for zero modes on the interval $0\le x<\infty$, which appear when the topological invariant at long distances $W_{R}=W(x\to\infty)$ is nonzero.
In the case that the local topological invariant $W(x)$ is constant on the interval $0\le x<\infty$, then the zero modes will localize at $x=0$.
In the case where $w\gg\xi$ however, if the local topological invariant $W(x)$ changes at the points $x_i^*$ with $|x_i^*|\lesssim w$ and the distance between the points is larger than the localization length $\xi$, then the zero modes will localize at the points $x\approx x_i$, as expected again from the bulk-boundary correspondence.

Hence, the local topological invariant encodes the information about the localization of the zero modes, which may not localize near the center of the smooth domain wall if $w\gg \xi$.
On the other hand, we have seen that the quantity $ v_{L,R}^2+ m_{L,R}$ determines whether the zero modes exhibit exponentially damped oscillations ($ v_{L,R}^2+ m_{L,R}<0$) or a simple exponential decay without oscillations ($ v_{L,R}^2+ m_{L,R}\ge0$).
Similarly, one can surmise that the sign of the function $ v(x)^2+ m(x)$ will indicate the presence of oscillations for $ v(x)^2+ m(x)<0$ and the absence of oscillations for $ v(x)^2+ m(x)\ge0$ at any given point $x$.
In particular, it is possible that the rate of the oscillations may change considerably around the center of the domain wall, i.e., in the region $|x| \lesssim w$.
More remarkably, oscillations may appear only within an interval $|x| \lesssim x_\text{osc}$ if $ v(x)^2+ m(x)<0$ for $|x| < x_\text{osc}$ and $ v(x)^2+ m(x)= v_{L,R}^2+ m_{L,R}\ge0$ for $|x| >x_\text{osc}$ or conversely, only
for $|x| > x_\text{osc}$ if $ v(x)^2+ m(x)\ge0$ for $|x| < x_\text{osc}$ and $ v(x)^2+ m(x)= v_{L,R}^2+ m_{L,R}<0$ for $|x| >x_\text{osc}$.

These effects may be quite remarkable in the case of long-hair zero modes $w\gg\xi$ when the field around the domain wall is considerably large $| m_D|\gg| m_{L,R}|$ (e.g., in the case of symmetric or asymmetric Pöschl–Teller potential).
In this case, the zero mode may localize at $x\approx x_{1,2}^*$ while oscillations may appear only for $|x|< |x_\text{osc}|$ or far away $|x|> |x_\text{osc}|$ with $|x_\text{osc}|\gg \xi$.

\Cref{fig:infinfshort} show zero modes on the interval $-\infty<x<\infty$ in the case where the width of the smooth domain wall $w$ is smaller than or comparable with the characteristic decay lengths $\xi$, for different choices of the parameters $ m_{0,1,2}$ and $ v_{1,2}$, plotted using  \cref{eq:2generalsolutionsMAIN,eq:2generalsolutionshyperMAIN}.
These figures show examples of modes with exponential decay or exponentially damped oscillations.
The exponential decay of the modes is clearly visible, with the exception of a small region for $|x|<w<\xi$:
the modes have "short-hair", i.e., they are nonfeatureless for $|x|\lesssim w$ but can be regarded as featureless for distances $|x|\gtrsim w$.
\Cref{fig:infinflong} show cases analogous to \cref{fig:infinfshort}, but where the width of the smooth domain wall $w$ is larger than the characteristic decay lengths $\xi$.
The most noticeable difference with the previous case is that the exponential decay of the modes is now visible only at large distances $|x|\gtrsim w>\xi$:
the modes have "long-hair", i.e., they are nonfeatureless at finite length scales.

The local features of the zero modes become prominent when $w\gg\xi$ (long-hair),
especially when the field $m(x)$ shows large variations around the domain wall, i.e., when the component $m_2$ of the field increases.
This is particularly visible in \cref{fig:infinflong}(d) where $w>\xi$, 
with oscillations visible in a wide region close to the center where $ v(x)^2+ m(x)<0$ but disappear at larger distances where $ v(x)^2+ m(x)>0$,
and to some extent in \cref{fig:infinfshort}(d) where $w<\xi$, with oscillations visible in a narrower region.
Conversely, in \cref{fig:infinflong}(f), oscillations seem to occur only at large distances where $ v(x)^2+ m(x)<0$ but disappear at smaller distances where $ v(x)^2+ m(x)>0$. 
Even more remarkably, the zero mode in \cref{fig:infinflong}(d) is localized far away from the center of the domain wall in correspondence of a point $|x_1^*|>\xi$, while  the two zero modes in \cref{fig:infinflong}(f) are localized around the two points $|x_{1,2}^*|>\xi$ where the local topological invariant $W(x)$ changes and the field $ m(x)$ changes its sign.

\section*{Experimental detection}

The definition of the decay rates and momenta mandates that $\mu_{L}^\pm \pm{\ii} \kappa_{L}=-s v_L\pm\sqrt{ v_{L}^2+ m_{L}}$ and $\mu_{R}^\pm \pm{\ii} \kappa_{R}=+s v_R\pm\sqrt{ v_{R}^2+ m_{R}}$, which we can summarize as
\begin{equation}\label{eq:EXP1}
\mu_{L,R}\pm{\ii} \kappa_{L,R}
=\mp v_{L,R}\pm\sqrt{ v_{L,R}^2+ m_{L,R}},
\end{equation}
which is valid for both featureless and nonfeatureless zero modes.
We notice that all the quantities on the left hand of these equations describe the wavefunction of the modes at large distances.
In condensed matter systems, the wavefunctions of such topologically zero modes are in principle measurable by spatially-resolved spectroscopies, while the quantities on the right hand of these equations depend on the "bulk" properties of the system at large distances.
In particular, we notice that $ m_{L,R}$ corresponds to the gap while the term $ v_{L,R}$ to the Fermi velocity of the gapped Dirac cone describing the dispersion of bulk excitations of the system at large distances $x/w\to\pm\infty$ on the left and on the right of the domain wall.
Hence, it is possible, in principle, to design an experiment in which some external parameters are varied and, for each setup, measure the dispersion of the bulk excitations at large distances and the decay rate and momenta of the zero modes simultaneously.
From these two sets of independent measures, one can thus obtain the quantities $ m_{L,R}^\text{exp}$, $ v_{L,R}^\text{exp}$ (bulk excitations) and $\mu_{L,R}^\text{exp}$, $\kappa_{L,R}^\text{exp}$ (zero modes).
Strong evidence of the presence of zero modes is obtained when these quantities satisfy \cref{eq:EXP1} on an extended range of external parameters.

If the modes exhibit exponentially damped oscillations $\kappa_{L,R}\neq0$, one also obtains that $\mu_{L,R}=\mp v_{L,R}$ and consequently
\begin{equation}\label{eq:EXP2}
\kappa_{L,R}^2+\mu_{L,R}^2
=- m_{L,R},
\end{equation}
which is valid for both featureless and nonfeatureless zero modes.
The left hand of this equation depends on the asymptotic behavior of the modes at large distances, while the right hand depends only on the bulk gap $ m_{L,R}$ and not on the Fermi velocity at large distances $x/w\to\pm\infty$.
By varying the external parameters, one can measure the bulk gap at large distances and, simultaneously, the decay rate and momenta of the zero modes.
Strong evidence of the presence of zero modes is obtained when the measured quantities $ m_{L,R}^\text{exp}$ (bulk gap) and $\mu_{L,R}^\text{exp}$, $\kappa_{L,R}^\text{exp}$ (zero modes) satisfy \cref{eq:EXP2} on an extended range of external parameters.

\section*{Internal symmetry and duality}

The Sturm-Liouville form of the modified Jackiw-Rebbi equation in \cref{eq:diffeq} is
\begin{equation}\label{eq:SL}
\partial_x
\left(
{\ee}^{2s\int  v(x) {\dd} x}
\varphi'(x)
\right)
-
{\ee}^{2s\int  v(x) {\dd} x}
 m(x)\varphi(x)
=0.
\end{equation}
The modified Jackiw-Rebbi equation can also be transformed into 
\begin{equation}\label{eq:SchrodingerGeneral}
\widetilde\varphi''(x)
-
\left( v(x)^2+s v'(x)+ m(x)\right)
\widetilde\varphi(x)
=0,
\quad
\widetilde\varphi(x)={\ee}^{s {\int v(x){\dd} x}}\varphi(x),
\end{equation}
which is again in the form of \cref{eq:diffeq} when $ v(x)\to0$ and $ m(x)\to m(x)+{ v(x)}^2+s v'(x)$, but and also equivalent to the Schrödinger equation of a particle with energy $E$ scattering off a potential $ V(x)$ with $ V(x)-E= v(x)^2+s v'(x)+ m(x)$, given by
\begin{equation}\label{eq:Schrodinger}
\left(
\eta
p^2
+
 V(x)
\right)
\widetilde\varphi(x)
=E \widetilde\varphi(x),
\end{equation}
where we take $\eta=1/2m=1$.
In the simple case of a uniform field $ v(x)= v$, the modified Jackiw-Rebbi equation is transformed into 
\begin{equation}\label{eq:SchrodingerUniform}
\widetilde\varphi''(x)
-
\left( v^2+ m(x)\right)
\widetilde\varphi(x)
=0,
\quad
\widetilde\varphi(x)={\ee}^{s v x}\varphi(x),
\end{equation}
which is in the form of \cref{eq:diffeq} when $ v(x)\to0$ and $ m(x)\to m(x)+ v^2$, and equivalent to the Schrödinger equation of a particle with energy $E=- v^2$ scattering off a potential $ V(x)= m(x)$, or alternatively, in a potential $ V(x)= v^2+ m(x)$ at zero energy.

Therefore, there is a one-to-one correspondence between general solutions of the modified Jackiw-Rebbi equation and the Schrödinger equation described by \cref{eq:SchrodingerGeneral}.
We will show below that this also induces a one-to-one correspondence between the particular solutions of these equations, in particular between the particular solutions of the modified Jackiw-Rebbi equation satisfying the boundary conditions $\varphi(x\to\pm\infty)=0$ (or $\varphi(x\to\pm\infty)=\varphi(0)=0$) and the particular solutions of the associated Schrödinger equation satisfying the boundary conditions $|\widetilde\varphi(x\to\pm\infty)|<\infty$ on the infinite interval $-\infty<x<\infty$ (or $|\widetilde\varphi(x\to\infty)|<\infty$, $\widetilde\varphi(0)=0$ on the semi-infinite interval $0\le x<\infty$).
This correspondence induces a duality between the theory of topological insulators and superconductors described by the modified Jackiw-Rebbi equation, and the theory of spinless nonrelativistic particles described by the Schrödinger equation.

We will now show that the transformation in \cref{eq:SchrodingerGeneral} implies the existence of an internal symmetry in the space of all possible modified Jackiw-Rebbi equations in \cref{eq:H-pwaveC} spanned by the possible choices of the fields $ m(x)$ and $ v(x)$.
This space includes as a special case the Schrödinger equation in \cref{eq:Schrodinger} obtained when $ v(x)=0$.
Let us take $\varphi\to\varphi_{\rm I}$ and $\varphi\to\varphi_{\rm II}$ in \cref{eq:diffeq} corresponding to the fields $ m_{\rm I}, v_{\rm I}$ and $ m_{\rm II}, v_{\rm II}$ giving respectively two distinct modified Jackiw-Rebbi equations
$\varphi_{\rm I}''(x)
+{2s v_{\rm I}}(x) \varphi_{\rm I}'(x)
-
 m_{\rm I}(x)
\varphi_{\rm I}(x)
=0$
and
$\varphi_{\rm II}''(x)
+{2s v_{\rm II}}(x) \varphi_{\rm II}'(x)
-
 m_{\rm II}(x)
\varphi_{\rm II}(x)
=0$.
If 
\begin{equation}\label{eq:invariantfields}
 m_{\rm I}(x)+ v_{\rm I}(x)^2+s v_{\rm I}'(x)=
 m_{\rm II}(x)+ v_{\rm II}(x)^2+s v_{\rm II}'(x),
\end{equation}
then from \cref{eq:SchrodingerGeneral} directly follows that both modified Jackiw-Rebbi equations map into an identical Schrödinger equation in the form of \cref{eq:SchrodingerGeneral}, and therefore their solutions are given by $\varphi_{\rm I,II}(x)=\widetilde\varphi(x){\ee}^{-s {\int v_{\rm I,II}(x){\dd} x}}$, which also mandates that
\begin{equation}
\varphi_{\rm I}(x){\ee}^{s {\int v_{\rm I}(x){\dd} x}}
=
\varphi_{\rm II}(x){\ee}^{s {\int v_{\rm II}(x){\dd} x}},
\end{equation}
or alternatively
\begin{equation}
\varphi_{\rm I}(x)
=
\varphi_{\rm II}(x){\ee}^{s {\int\left( v_{\rm II}(x)- v_{\rm I}(x)\right){\dd} x}}.
\end{equation}
This property induces an equivalence relation or internal symmetry in the space of all possible equations in the form \cref{eq:H-pwaveC} which leaves $ v(x)^2+s v'(x)+ m(x)$ invariant.
A special case is obtained when one of the two equations has uniform field $ v(x)= v$, which gives
$ m_{\rm I}(x) +  v_{\rm I}^2=
 m_{\rm II}(x) +  v_{\rm II}(x)^2+s v_{\rm II}'(x)$,
and when
$ v(x)=0$ (i.e., the modified Jackiw-Rebbi reduces to a simpler Schrödinger equation), which gives
$ m_{\rm I}(x)= m_{\rm II}(x) +  v_{\rm II}(x)^2+s v_{\rm II}'(x)$.
Moreover, since $ v'(x)=0$ for $x\to\pm\infty$, 
the quantities $ v_{L,R}^2+ m_{L,R}$ and $ q_{L,R}$
are invariant under the internal symmetry, i.e., 
$( v_{L,R}^2+ m_{L,R})_{\rm I}=( v_{L,R}^2+ m_{L,R})_{\rm II}$ and
$( q_{L,R})_{\rm I}=( q_{L,R})_{\rm II}$.
However, the asymptotic values of the fields $ m_{L,R}$ and $ v_{L,R}$ are not invariant under the internal symmetry, neither are the exponents $\alpha_{L,R}^\pm$, $\mu_{L,R}^\pm$ and the topological invariants $W_{L,R}$.
The conservation of the quantities $ v_{L,R}^2+ m_{L,R}$ mandates that the oscillatory character of the solutions is conserved under the internal symmetry.
In the limiting case $ v_{L,R}=0$, however, all exponents $\alpha_{L,R}^\pm$, $\mu_{L,R}^\pm$, $\kappa_{L,R}$, the quantities $ m_{L,R}$ and trivially the topological invariants $W_{L,R}=0$ are invariant under the internal symmetry.
This limiting case is not very interesting because no zero modes can exist for $ v_{L,R}=0$ (the boundary conditions cannot be satisfied in this case).

It is important to notice that since the existence and number of zero modes crucially depend on the topological invariants $W_{L,R}$, and 
since the internal symmetry does not preserve these quantities (with the exception of the limiting case $ v_{L,R}=0$), the existence and number of zero modes is not conserved by the internal symmetry.
In other words, the internal symmetry does not preserve the existence of particular solutions that satisfy the boundary conditions.

In the case where the fields $ v(x)$ and $ m(x)$ can be expanded in powers of $y(x)$ up to the first and second order, then the integral $\int v(x){\dd} x$ in \cref{eq:SchrodingerGeneral} simplifies giving
\begin{equation}\label{eq:SchrodingerGeneralIntegral}
\widetilde\varphi(x)=
\left(2\cosh\left(\frac{x}{2w}\right)\right)^{s w v_1 }{\ee}^{s (2v_0 + v_1) x/2}\varphi(x)
,
\end{equation}
where we fixed the arbitrary prefactor to be consistent with \cref{eq:2generalsolutionsMAIN}.
Moreover, one gets
$ v(x)^2+s v'(x)+ m(x)=
( v_0^2+ m_0)+
(2 v_0 v_1+(s/w) v_1+ m_1)y(x)+
( v_1^2-(s/w) v_1+ m_2)y(x)^2$
such that the condition \cref{eq:invariantfields} becomes
\begin{subequations}
\begin{align}
( v_0^2+ m_0)_{\rm I}&=( v_0^2+ m_0)_{\rm II},
\\
(2 v_0 v_1+(s/w) v_1+ m_1)_{\rm I}&=(2 v_0 v_1+(s/w) v_1+ m_1)_{\rm II},
\\
( v_1^2-(s/w) v_1+ m_2)_{\rm I}&=( v_1^2-(s/w) v_1+ m_2)_{\rm II},
\end{align}
\end{subequations}
which means that the quantities
$ v_0^2+ m_0$,
$2 v_0 v_1+(s/w) v_1+ m_1$, and
$ v_1^2-(s/w) v_1+ m_2$
are invariant under the internal symmetry.
These equations, together with the fact that $( q_{L,R})_{\rm I}=( q_{L,R})_{\rm II}$, also mandate that the values of the parameters $a_{1,2}$, $b_{1,2}$, $c_{1,2}$ are invariant under the internal symmetry, i.e.,
$(a_{1,2})_{\rm I}=(a_{1,2})_{\rm II}$, 
$(b_{1,2})_{\rm I}=(b_{1,2})_{\rm II}$, and
$(c_{1,2})_{\rm I}=(c_{1,2})_{\rm II}$.
A special case is obtained when one of the two equations has uniform field $ v(x)= v= v_0$, which gives
$( v_0^2+ m_0)_{\rm I}=( v_0^2+ m_0)_{\rm II}$,
$( m_1)_{\rm I}=(2 v_0 v_1+(s/w) v_1+ m_1)_{\rm II}$,
and
$( m_2)_{\rm I}=( v_1^2-(s/w) v_1+ m_2)_{\rm II}$.

\section*{Bounded modes}

The solutions of the associated Schrödinger equation in \cref{eq:Schrodinger} are closely related to the solutions of the modified Jackiw-Rebbi equation in \cref{eq:diffeq}, and can be obtained from the previous results by 
substituting $ v(x)\to0$ and $ m(x)\to V(x)-E$.
Assuming that the potential approaches its limit values on the left and right with exponential decay at large distances $|x|\gg w$ as
$| V(x\to\mp\infty)- V_{L,R}|\propto{\ee}^{-|x|/w}$, where $w>0$ is a characteristic length describing the spatial variations of the potential, the general solution of the associated Schrödinger equation is given by a linear combination $\sum_i A_i\widetilde\varphi_i(x)$ where
\begin{align}
 \widetilde\varphi_{1,2}(x)=&
 \frac{{\ee}^{\pm({  q_L}+{ q_R})x/2} }
 {\left(2\cosh{\left(\frac{x}{2w}\right)}\right)^{{\pm w( q_L}-{ q_R})}}
 F_{1,2}(x)
 ,
\label{eq:2generalsolutionsSCMAIN}
\end{align}
with $ q_{L,R}=\sqrt{ V_{L,R}-E}$
depending only on the values of the potential $ V_{L,R}$ at large distances $|x|>w$.
Asymptotically, these solutions give 
$\widetilde\varphi_{1,2}(x\to-\infty)\propto{\ee}^{\pm  q_L x}$,
and $\widetilde\varphi_{1,2}(x\to+\infty)\propto{\ee}^{\pm  q_R x}$.
For each side, one has either 
$\Re( q_{L,R})=\mu_{L,R}=1/\xi_{L,R}\neq0$ and $\Im( q_{L,R})=0$ for $ V_{L,R}-E>0$ or 
$\Im( q_{L,R})=\kappa_{L,R}=2\pi/\lambda_{L,R}\neq0$ and $\Re( q_{L,R})=0$ for $ V_{L,R}-E<0$, and $ q_{L,R}=0$ in the limiting case $ V_{L,R}=E$.
If the potential can be expanded as $ V(x)= V_0+ V_1 y(x)+ V_2 y(x)^2$, the general solution is given using $F_{1,2}$ as in 
\cref{eq:2generalsolutionshyperMAIN} with parameters given by \cref{eq:ab1ab2MAIN} 
taking $ v_1\to0$ and $ m_2\to  V_2$,
with the exception of the case where 
$ V(x)=E$, where the only bounded solution is $\widetilde\varphi(x)=\text{const}$.
The solutions $\widetilde\varphi_{1,2}(x)$ simplify in the cases where $ V_{L,R}= V$ or if the potential are uniform, as in \cref{tab:specialcasesSC}.
Again, the exponents $\pm q_L$ and $\pm q_R$ uniquely determine the asymptotic behavior of the solutions and whether the solutions can satisfy the boundary conditions on a given interval.

\begin{table*}[t]
 \centering
 \includegraphics[width=\textwidth]{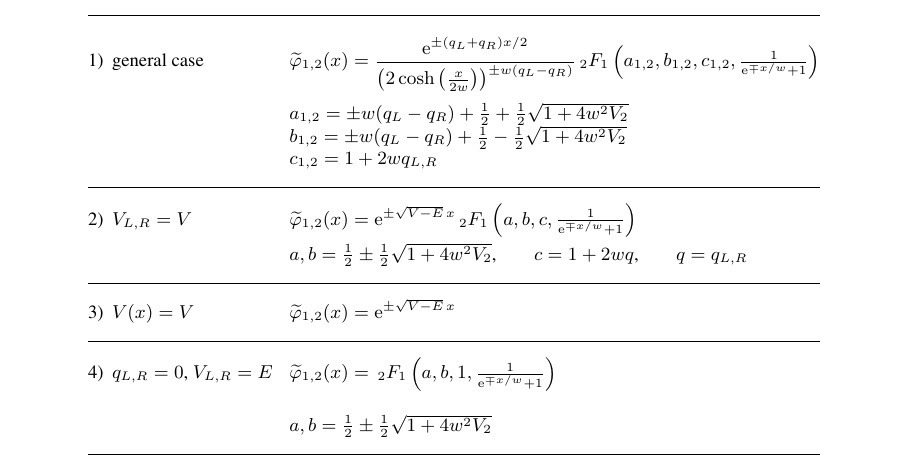} 
\caption{
The general solutions of the associated Schrödinger equation in terms of hypergeometric functions
for 
$ V(x)= V_0+ V_1 y(x)+ V_2 y(x)^2$ 
 and 
for some special cases, i.e.,
where $ V(x)= V$ is constant
or a symmetric Pöschl–Teller potential 
($ V_0= V$ and $ V_1+ V_2=0$), 
with $ V_{L,R}= V$.
If $ q_{L,R}= q$, one gets
$a_{1,2}=a$ and $b_{1,2}=b$ with $a+b=1$ in rows 2 and 4:
In this case
one has also
  ${{\,}_2F_1\left(a,b,c,\frac{1}{{\ee}^{\mp x/w}+1}\right)}
=\Gamma(c)
{\ee}^{\pm  q x} P^{1-c}_{-a}\left(\tanh\left(\mp\frac{x}{2w}\right)\right)
$.
}
 \label{tab:specialcasesSC}
\end{table*}

In the case of the Schrödinger equation as well, the values of the fields at large distances and the boundary conditions univocally determine the existence and number of particular solutions on infinite and semi-infinite intervals.
We define
\begin{equation}\label{eq:TISC}
\widetilde W_{L,R}=
\begin{cases}
1 & \text{ if }  V_{L,R}-E\le0,
\\
0 & \text{ if }  V_{L,R}-E>0.
\end{cases}
\end{equation}
Bounded modes on the interval $-\infty<x<\infty$ are the modes $\widetilde\varphi_{1,2}(x)$ which satisfy the boundary conditions $|\widetilde\varphi(x\to\pm\infty)|<\infty$.
If $ V_{R}-E\le0$, the mode $\widetilde\varphi_1(x)$ satisfies the boundary conditions.
On the other hand, if $ V_{L}-E\le0$, the mode $\widetilde\varphi_2(x)$ satisfies the boundary conditions.
Hence, we get two modes for $\widetilde W_L=\widetilde W_R=1$, one mode for $\widetilde W_L\neq\widetilde W_R$, and no modes for $\widetilde W_L=\widetilde W_R=0$.
If $\widetilde W_L=\widetilde W_R=1$, i.e., if $ V_{L,R}\le E$, we get two linearly independent modes $\widetilde\varphi_{1,2}$ showing oscillatory behavior with momentum $\kappa_{L,R}=\Im( q_{L,R})$ on the left and on the right asymptote $x\to\pm\infty$, respectively. 
If $\widetilde W_R=1$ and $\widetilde W_L=0$, i.e., 
if $ V_L>E\ge V_{R}$,
we get only one mode $\widetilde\varphi_{1}$.
On the right side, this mode shows oscillatory behavior with momentum $\kappa_{R}=\Im( q_{R})$.
On the left side, the mode shows exponential decay with $\mu_{L}= q_{L}$.
If $\widetilde W_R=0$ and $\widetilde W_L=1$, i.e., 
if $ V_R>E\ge V_L$, we get only one mode $\widetilde\varphi_{2}$, with analogous behavior on the left and right, respectively.

Bounded modes on the interval $0\le x<\infty$ are the modes $\widetilde\varphi_{1,2}(x)$ which satisfy the boundary conditions $\widetilde\varphi(0)=0$ and $|\widetilde\varphi(x\to\infty)|<\infty$.
If $\widetilde W_R=1$, i.e., if $E\ge V_R$, 
both $\widetilde\varphi_{1,2}(x)$ satisfy the boundary conditions for $x\to\infty$.
Hence in this case, a particular solution is the linear combination $\widetilde\varphi(x)=\sum_i A_i\widetilde\varphi_i(x)$ with $A_{1,2}$ chosen such that $\widetilde\varphi(0)=0$.
This mode shows oscillatory behavior with momentum $\kappa_{R}=\Im( q_{R})$.
In the case where $ V_R= V_L$, one has ${A_1}/{A_2} = -1$.
Moreover, if the potential is constant $ V(x)=0$, we recover the well-known textbook solution $\varphi(x)=\sin(\kappa x)$ describing the reflection of a free particle with momentum $\kappa=\sqrt{E- V}$ scattering off an infinite wall. 
Bounded modes on the interval $-\infty<x\le 0$ can be obtained similarly.
The existence and number of bounded modes on the infinite and finite intervals $-\infty<x<\infty$ and $0\le x<\infty$ is summarized in \cref{tab:solSCintervals}.

Notice that we required different kinds of boundary conditions for the zero-energy solutions of the generalized Jackiw-Rebbi equation and for the associated Schrödinger equation.
In the case of the generalized Jackiw-Rebbi equation, we are concerned about normalizable modes that satisfy the Dirichlet boundary conditions $\varphi(\pm\infty)=0$.
These normalizable modes do not exist, in general, for the associated Schrödinger equation.
In this case, we only require that the modes are bounded $|\widetilde\varphi(x)|<\infty$ for $x\to\pm\infty$. 
At long distances, these modes behave either as plane waves or decay exponentially.

\section*{Wavelike modes and Shockley-like modes}

Hence, the value of the potential at large distances and the boundary conditions univocally determine the existence and number of bounded modes on infinite and semi-infinite intervals, in particular, on the values of the potential $ V_{L,R}$ for modes on the infinite interval $-\infty<x<\infty$, and on the value of the potential $ V_R$ on the semi-infinite interval $0\le x<\infty$.

\begin{table}[t]
 \centering
 \includegraphics[width=\textwidth]{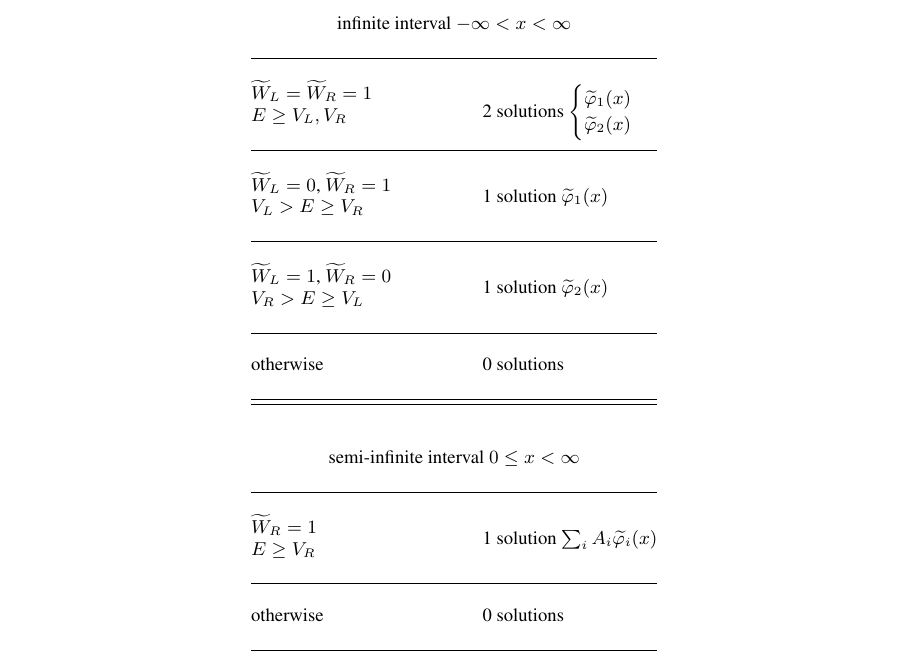} 
\caption{
The number of solutions of the associated Schrödinger equation satisfying the boundary conditions 
$\widetilde\varphi(\pm\infty)<\infty$ on the infinite interval $-\infty<x<\infty$ and
$\widetilde\varphi(0)=0$, $\widetilde\varphi(\infty)<\infty$ on the semi-infinite interval $0\le x<\infty$ 
depending on the energy $E$ and on the asymptotic values of the potential $ V_{L,R}$.
}
 \label{tab:solSCintervals}
\end{table}

The bounded modes $\widetilde\varphi_{1,2}(x)$ on the infinite interval $-\infty<x<\infty$ with $\widetilde W_L=\widetilde W_R=1$, i.e., $E\ge V_{L,R}$, describe the transmission and reflection of particles with energies greater than the height of the domain wall $E\ge\max( V_{L,R})$.
These modes are wavelike modes with oscillatory having momenta $\kappa_{L,R}=\Im( q_{L,R})=\sqrt{E- V_{L,R}}$, respectively on each side of the domain wall.
Hence, these modes are completely delocalized.
When $w\to0$ and $ V(x)= V_L= V_R$, this simply reduces to the case of a free particle with momentum $\kappa=\sqrt{E- V}$.

The bounded mode $\widetilde\varphi_{1}(x)$ with $\widetilde W_L=0$ and $\widetilde W_R=1$, i.e., with $ V_{L}>E\ge V_R$,
describes instead the tunneling and reflection of a particle with energy smaller than the wall height $E<\max( V_{L,R})$.
On the right side of the domain wall, the mode shows oscillatory behavior with momentum $\kappa_R=\sqrt{E- V_{R}}$.
On the left side of the smooth domain wall instead, it decays exponentially with decay rate $\mu_L=\sqrt{ V_{L}-E}$.
A finite decay rate mandates a finite tunneling probability to find the particle beyond the domain wall.
We refer to this mode, which is constant or oscillating on one side of the domain wall and exponentially decaying on the other, as a Shockley-like mode.
Indeed, this mode closely resembles a Shockley state, i.e., an electronic state at the surface of a metal or a semiconductor, which exhibits an exponentially decaying tail outside the surface.
Indeed, when $w\to0$, this case yields the well-known textbook solution of a particle scattering off a step potential, which coincides exactly with the wavefunction of a Shockley state localized at the surface of a metal or a semiconductor when the potential inside the bulk material is constant.

Moreover, the bounded mode on the semi-infinite interval $0\le x<\infty$ with $\widetilde W_R=1$, i.e., $E\ge V_{R}$ describes the reflection of a particle with momentum $\kappa=\sqrt{E- V_R}$ off an infinite wall with smooth spatial variations of the potential near $x=0$, which for $w\to0$ reduces to the well-known case describing the reflection of a free particle scattering off an infinite wall.

Again, the asymptotic behavior at large distances $x/w\to\pm\infty$ is determined by the characteristic decay lengths $\xi_{L,R}$ and wavelengths $\lambda_{L,R}$ for each of the left and right sides, which ultimately depend only on the values of the potential at the left and right asymptotes $ V_{L,R}$.
At short distances $|x|\lesssim w$ instead, the wavefunctions do not only depend on the values of the potential at large distances $ V_{L,R}$, but also by the width of the smooth domain wall $w$, and on the details of the spatial dependences of the potential.

As we anticipated, the duality between the general solutions of the modified Jackiw-Rebbi equation and the Schrödinger equation described by \cref{eq:SchrodingerGeneral} induces a duality between the particular solutions of these equations on the infinite interval $-\infty<x<\infty$ satisfying the boundary conditions $\varphi(x\to\pm\infty)=0$ and $|\widetilde\varphi(x\to\pm\infty)|<\infty$, respectively, and between the particular solutions of these equations on the semi-infinite interval $0\le x<\infty$ satisfying
the boundary conditions $\varphi(x\to\pm\infty)=\varphi(0)=0$ and $|\widetilde\varphi(x\to\infty)|<\infty$, $\widetilde\varphi(0)=0$, respectively.

We consider the duality transformation in \cref{eq:SchrodingerGeneral}, giving $ v(x)^2+s v'(x)+ m(x)= V(x)-E$.
This transformation mandates that, if $\widetilde W_{L,R}=1$ then $|W_{L,R}|=1$ assuming $ v_{L,R}\neq0$, since in this case $ v_{L,R}^2+ m_{L,R}=V_{L,R}-E\le0$.
The converse is not always true, since $|W_{L,R}|=1$ mandates $\widetilde W_{L,R}=1$ only if $| m_{L,R}|> v_{L,R}^2$.
Hence, the 2 bounded modes $\widetilde\varphi_{1,2}(x)$ on the infinite interval $-\infty<x<\infty$ with $\widetilde W_L=\widetilde W_R=1$, i.e., 
with $E\ge V_{L,R}$ correspond via the duality transformation to the 2 zero modes $\varphi_{1,2}^s(x)$, if we assume $ v_L v_R<0$, since in this case $|W_{L,R}|=1$ and $W_L W_R=-1$.
The two bounded modes and the zero modes share the same oscillatory character, with the two bounded modes oscillating and the two zero modes decaying exponentially with damped oscillations and decay rates $| v_{L,R}|$.
On each side of the barrier, these oscillations have the same momenta, which are given by $\kappa_{L,R}=\Im(\sqrt{ v_{L,R}^2+ m_{L,R}})=\Im(\sqrt{V_{L,R}-E})$.
In the limiting cases where either of these momenta vanishes, the bounded modes converge to a constant value and the zero modes decay exponentially without oscillations, respectively on the left (if $E= V_{L}$), on the right (if $E= V_{R}$), and on both sides (if $E= V_L= V_R$).

The bounded mode $\widetilde\varphi_{1}(x)$ with $\widetilde W_L=0$ and $\widetilde W_R=1$ with $ V_{L}>E\ge V_R$ correspond via the duality transformation to the zero mode $\varphi_{1}^s(x)$ with $W_{L}=0$ and $|W_{R}|=1$ if we assume that $| m_{L}|> v_{L}^2$ and $ v_R\neq0$.
On the right side, the bounded mode is oscillating and the zero node is exponentially decaying with damped oscillations and decay rate $| v_{R}|$.
The oscillations have same momentum $\kappa_{R}=\Im(\sqrt{ v_{R}^2+ m_{R}})=\Im(\sqrt{V_{R}-E})$.
Again, in the limiting cases where this momentum vanishes (i.e., if $E= V_{R}$) the bounded mode converges to a constant and the zero mode decays exponentially without oscillations.
On the left side, the bounded mode and zero mode decay exponentially with decay rates $\mu_L= q_L$ and $\mu_L^+=-s v_L\pm q_L$, respectively.
Same considerations can be made for the bounded mode $\widetilde\varphi_{2}(x)$ and zero mode $\varphi_{2}^s(x)$.

Finally, the bounded mode on the semi-infinite interval $0\le x<\infty$
with $\widetilde W_R=1$, i.e., $E\ge V_{R}$, corresponds via the duality transformation to the zero mode if we assume $ v_R\neq0$.
The bounded mode is oscillating and the zero node is exponentially decaying with damped oscillations, with same momentum $\kappa_{R}=\Im(\sqrt{ v_{R}^2+ m_{R}})=\Im(\sqrt{V_{R}-E})$.
In the limiting case, $E= V_{R}$) the bounded mode converges to a constant and the zero node decays exponentially without oscillations.
The duality between the bounded modes and the zero modes is summarized in \cref{tab:duality}.

\begin{table}[t]
 \centering
 \includegraphics[width=\columnwidth]{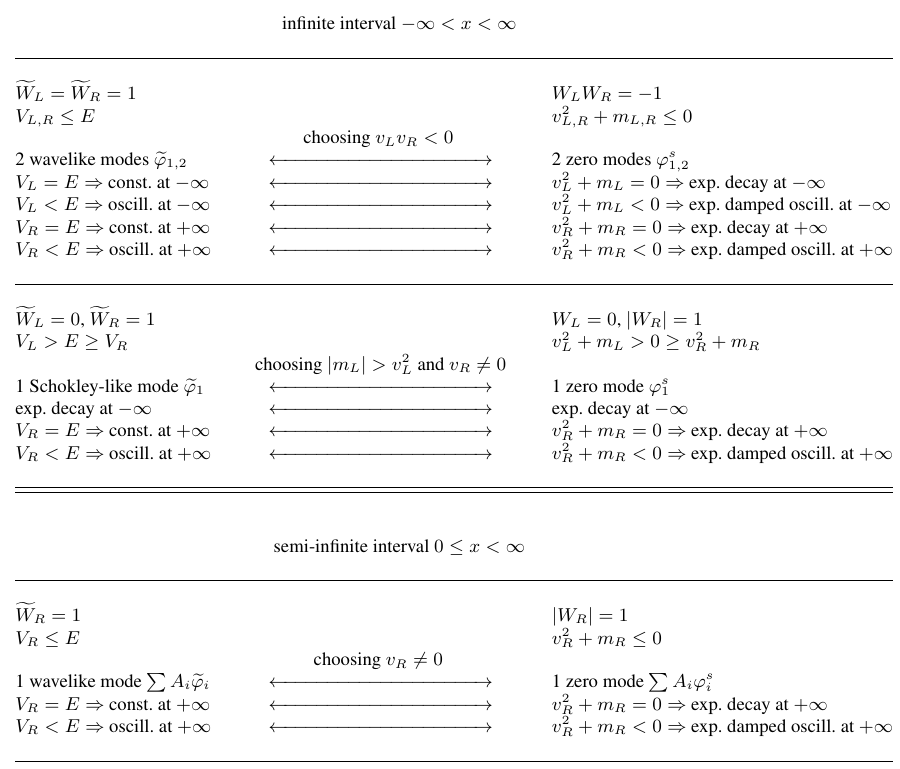} 
\caption{
Duality between the zero modes of the modified Jackiw-Rebbi equation and the bounded modes of a Schrödinger equation under the transformation $ v(x)^2+s v'(x)+ m(x)= V(x)-E$ on the infinite interval $-\infty<x<\infty$ and on the semi-infinite interval $0\le x<\infty$.
Notice that $\widetilde W_{L,R}=1$ mandates $|W_{L,R}|=1$ assuming $ v_{L,R}\neq0$
and that
$\widetilde W_{L,R}=0$ mandates $|W_{L,R}|=0$ assuming $| m_{L,R}|> v_{L,R}^2$ . 
Moreover one has $ v^2_{L,R}+ m_{L,R}=V_{L,R}-E$ in all cases.
Hence, there is a one-to-one correspondence between bounded modes and zero modes and between the oscillations of the bounded modes and damped oscillations of the zero modes.
}
 \label{tab:duality}
\end{table}

Hence, surprisingly, the zero modes of the modified Jackiw-Rebbi equation are directly related to the bounded modes of a Schrödinger equation of a particle scattering off a potential.
However, there are some important differences:
There are now no topological invariant, since the Hamiltonian is now topologically trivial for all parameters, no localized zero modes at the phase boundary, and consequently no bulk-boundary correspondence.
However, we can recover a sort of bulk-boundary correspondence governing the existence of zero modes at phase boundaries, if the role of the topological invariant is now played by quantity $\widetilde W$.
If the quantity $\widetilde W$ changes (i.e., if $ V-E$ changes sign) in correspondence of the smooth domain wall, there exists a localized Shockley-like mode which exponentially decays (or exhibits exponentially damped oscillations) on the side of the domain wall with $\widetilde W=0$ (the side with $ V-E>0$).
If the $\widetilde W$ does not change at the domain wall, there can exist only wavelike modes, or no modes at all.
Notice that this also forbid the existence of Shockley-like modes at the left end of the interval $0\le x<\infty$ if $\widetilde W_R=0$, assigning $\widetilde W_L=0$ to the "vacuum", i.e., by taking $ V(x)\to\infty$ for $x<0$ (i.e., $ V-E>0$).
This identification of $\widetilde W$ with a topological invariant and the consequent interpretation of localized Shockley-like modes in terms of a bulk-boundary correspondence is not based on topological considerations must be taken with a grain of salt.

Similarly to the case of the zero modes solutions of the modified Jackiw-Rebbi equation, to gain a physical intuition on the wavefunction of the bounded modes at short distances $x\lesssim w$, one can analyze the spatial dependence of $ V(x)-E$ and define the quantity
\begin{equation}\label{eq:TISCRS}
\widetilde W(x)=
\begin{cases}
1 & \text{ if }  V(x)-E\le0,
\\
0 & \text{ if }  V(x)-E>0,
\end{cases}
\end{equation}
which generalizes $\widetilde W_{L,R}$ defined in \cref{eq:TISC} to the whole interval $-\infty<x<\infty$. 
We have seen that the character of the bounded modes on the interval $-\infty<x<\infty$ at long distances on each side of the smooth boundary depends on the value of $\widetilde W_{L,R}$:
If $\widetilde W_{L,R}=1$, the mode has a wavelike character,
converging to a constant value or oscillating with no decay, while 
if $\widetilde W_{L,R}=0$ the mode has a Shockley-like character with exponential decay. 

The crossover between constant/oscillating behavior and decaying behavior occurs at $x/w\approx0$, i.e., for $|x_1|\lesssim w$, since the spatial variations of the fields are confined within the smooth domain wall $-w\lesssim x\lesssim w$.
Analogously to the case of zero modes, the points where this crossover occurs coincide with the points where the local topological invariant changes from 0 to 1.

\begin{figure}[tbp]
 \centering
 \includegraphics[width=.7\textwidth]{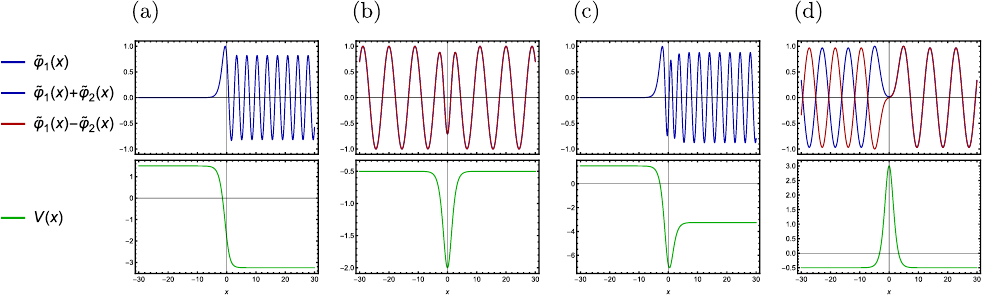}
\caption{
Bounded modes on the interval $-\infty<x<\infty$ with a smooth domain wall given by the field $ V(x)$ and $E=0$.
Upper panels show the wavefunction
and lower panels show the field $ V(x)$.
Different columns correspond via the duality transformation to the columns in \cref{fig:infinfshort}, by taking
$ V_0=( v_0^2+ m_0)$,
$ V_1=(2 v_0 v_1+(s/w) v_1+ m_1)$, and
$ V_2=( v_1^2-(s/w) v_1+ m_2)$
with $s=1$.
This corresponds to:
(a) 
single Shockley-like mode with exponential decay on the left side and oscillations on the right side
of a smooth domain wall with S-shaped field $ V(x)$ 
[corresponding to \cref{fig:infinfshort}(b)];
(b) 
two wavelike modes with oscillations on both sides
of a smooth domain wall with symmetric Pöschl–Teller field $ V(x)$ 
[corresponding to \cref{fig:infinfshort}(c) and taking the two linear combinations $\widetilde\varphi_1\pm\widetilde\varphi_2$ giving real wavefunctions];
(c) 
single Shockley-like mode with exponential decay on the left side and oscillations on the right side of a smooth domain wall with asymmetric Pöschl–Teller $ V(x)$ 
[corresponding to \cref{fig:infinfshort}(e)];
(d) 
two wavelike modes with oscillations on both sides
of a smooth domain wall with symmetric Pöschl–Teller field $ V(x)$ [corresponding to \cref{fig:infinfshort}(f) and taking the two linear combinations $\widetilde\varphi_1\pm\widetilde\varphi_2$].
}
 \label{fig:SCinfinfshort}
~\\
 \centering
 \includegraphics[width=.7\textwidth]{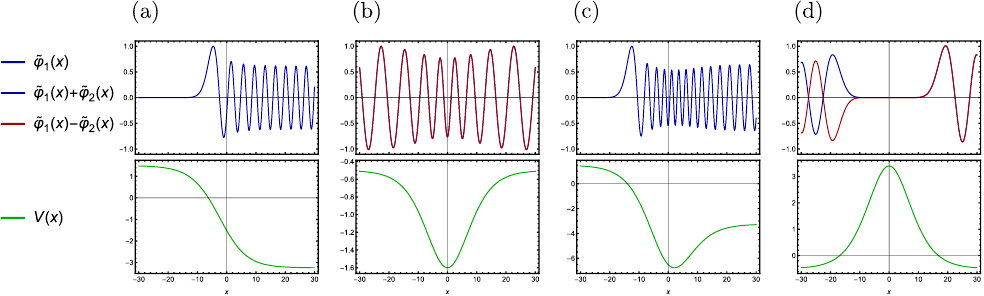} 
 \caption{
Bounded modes on the interval $-\infty<x<\infty$ as in \cref{fig:SCinfinfshort},
but where the width of the smooth domain wall is larger.
Different columns correspond via the duality transformation to the columns in \cref{fig:infinflong} with $s=1$.
In particular,
(a) 
single Shockley-like corresponding to \cref{fig:infinflong}(b),
(b) 
two wavelike modes corresponding to \cref{fig:infinflong}(c),
(c) 
single Shockley-like corresponding to \cref{fig:infinflong}(e),
(d) 
two wavelike modes corresponding to \cref{fig:infinflong}(f).
}
 \label{fig:SCinfinflong}
\end{figure}

\Cref{fig:SCinfinfshort} show bounded modes on the interval $-\infty<x<\infty$ for different choices of the parameters $ V_{0,1,2}$ and with $E=0$ plotted using \cref{eq:2generalsolutionsSCMAIN,eq:2generalsolutionshyperMAIN}.
\Cref{fig:SCinfinflong} shows the case where the width of the smooth domain wall $w$ becomes larger.
We choose the parameters such $ V_{0,1,2}$ to correspond to the parameters $ m_{0,1,2}$ and $ v_{1,2}$ via the duality transformations, i.e., taking 
$ V_0=( v_0^2+ m_0)$,
$ V_1=(2 v_0 v_1+(s/w) v_1+ m_1)$, and
$ V_2=( v_1^2-(s/w) v_1+ m_2)$
with $ m_{0,1,2}$ and $ v_{0,1}$ as in \cref{fig:infinfshort}(b), \cref{fig:infinfshort}(c),\cref{fig:infinfshort}(e), \cref{fig:infinfshort}(f), to obtain 
\cref{fig:SCinfinfshort}(a), \cref{fig:SCinfinfshort}(b), \cref{fig:SCinfinfshort}(c), \cref{fig:SCinfinfshort}(d), respectively,
and as in
\cref{fig:infinflong}(b), \cref{fig:infinflong}(c), \cref{fig:infinflong}(e), \cref{fig:infinflong}(f), to obtain 
\cref{fig:SCinfinflong}(a), \cref{fig:SCinfinflong}(b),\cref{fig:SCinfinflong}(c), \cref{fig:SCinfinflong}(d), respectively.
Analogously to the case of zero modes, the local features of the bounded modes become prominent when the width of the smooth domain wall increases.
This is especially visible in \cref{fig:SCinfinflong}(c), where the crossover between oscillations and decaying behavior occurs far away from the center of the domain wall, in correspondence with a point where the quantity $\widetilde W(x)$ changes, i.e., where the field $ V(x)$ changes its sign.

\section*{Conclusions}

In conclusion, we derived the analytical solutions of a modified Jackiw-Rebbi equation describing zero-energy modes at smooth domain walls between topologically trivial and nontrivial phases in one-dimensional topological insulators and superconductors. 
This allows for the characterizations of topologically-protected zero-modes in terms of a few length scales: 
the domain wall width, the decay length, and the oscillation wavelength of the wavefunction.
This provides a unified framework to describe physically different regimes, from featureless modes localized at sharp domain walls to nonfeatureless modes with "short" or "long" hair localized at smooth domain walls.
We also reveal a universal relation between the bulk gap, decay rate, and oscillation momentum of the zero modes, thus providing a quantifiable and experimentally measurable verification of the bulk-boundary correspondence. 
Additionally, we uncovered an unexpected duality between the topologically protected zero-modes and Shockley surface modes, highlighting a deeper connection between topological and nontopological states of matter.
These findings deepen the understanding of the localization properties of edge modes in topological insulators and Majorana zero modes in topological superconductors and shed some new light on the differences and similarities between topological and nontopological zero modes in these systems. 

\begin{acknowledgments} 
P.~M. thanks Junyong Eom and Krishanu Roy Chowdhury for the useful discussions.
%
P.~M. is supported by the Japan Science and Technology Agency (JST) of the Ministry of Education, Culture, Sports, Science and Technology (MEXT), JST CREST Grant~No.~JPMJCR19T2, the Japan Society for the Promotion of Science (JSPS) Grant-in-Aid for Early-Career Scientists KAKENHI Grant~No.~JP23K13028 and No.~JP20K14375.
\end{acknowledgments}

\end{document}